\documentstyle[epsfig,longtable]{aipproc}

\def\ale{\mathrel{\hbox{\rlap{\hbox{\lower4pt\hbox{$\sim$}}}\hbox{$<$}}}}
\def\age{\mathrel{\hbox{\rlap{\hbox{\lower4pt\hbox{$\sim$}}}\hbox{$>$}}}}

\begin{document}
\title{The afterglows of gamma-ray bursts}

\author{S. R. Kulkarni$^*$, E.~Berger$^*$, J.~S.~Bloom$^*$, 
  F.~Chaffee$^\P$, \\
  A.~Diercks$^*$,
  S.~G.~Djorgovski$^*$, D. A. Frail$^{\dagger}$,
  T.~J.~Galama$^*$, 
  R.~W.~Goodrich$^\P$ 
  F.~A.~Harrison$^*$, R.~Sari$^*$ \&\ S.~A.~Yost$^*$}

\address{$^*$California Institute of Technology, Pasadena, CA 91125,
    USA\\
$^\dagger$National Radio Astronomy Observatory,
      Socorro, NM 87801, USA \\
$^\P$W. M. Keck Observatory, Kamuela, HI 96743, USA}

\maketitle


\begin{abstract}
  Gamma-ray burst astronomy has undergone a revolution in the last
  three years, spurred by the discovery of fading long-wavelength
  counterparts.  We now know that at least the long duration GRBs lie at
  cosmological distances with estimated electromagnetic energy release
  of $10^{51}$ -- $10^{53}$ erg, making these the brightest explosions
  in the Universe.  In this article we review the current
  observational state, beginning with the statistics of X-ray,
  optical, and radio afterglow detections.  We then discuss the
  insights these observations have given to the progenitor population,
  the energetics of the GRB events, and the physics of the afterglow
  emission.  We focus particular attention on the evidence linking GRBs to
  the explosion of massive stars.  Throughout, we identify remaining
  puzzles and uncertainties, and emphasize promising observational
  tools for addressing them.  The imminent launch of {\em HETE-2}
  and the increasingly sophisticated and coordinated
  ground-based and space-based
  observations have primed this field for fantastic growth.

\end{abstract}

\section{Introduction}            
\label{sec:introduction}

GRBs have mystified and fascinated astronomers since their discovery.
Their brilliance and their short time
variability clearly suggest a compact object
(black hole or neutron star) origin. Three decades of high-energy
observations, culminating in the definitive measurements of
CGRO/BATSE, determined the spatial distribution to be isotropic yet
inhomogeneous, suggestive of an extragalactic population (see
\cite{fm95} for a review of the situation prior to the launch of the
BeppoSAX mission).  Further progress had to await the availability
of GRB positions adequate for identification of counterparts at
other wavelengths.

In the cosmological scenario, GRBs would have energy releases comparable
to that of supernovae (SNe). Based on this analogy,
Paczy\'nsk \&\ Rhoads \cite{pr93} and Katz \cite{k94}
predicted that the gamma-ray burst would be followed
by long-lived but fading emission.
These papers motivated systematic searches
for radio afterglow, including our effort at the VLA \cite{fk95}.
The broad-band nature of this
``afterglow'' and its detectability was underscored in later work
\cite{mr97,v97}. 

Ultimately, the detection of the predicted afterglow had to await
localizations provided by the Italian-Dutch satellite, BeppoSAX
\cite{b+97}.  The BeppoSAX Wide Field Camera (WFC) observes
about 3\% of the sky, triggering on the low-energy (2 -- 30 keV)
portion of the GRB spectrum, localizing events to $\sim$ 5 -- 10
arcminutes.  X-ray afterglow was first discovered by BeppoSAX in GRB~970228,
after the satellite was re-oriented (within
about 8 hours) to study the error circle of a WFC detection with the
2 -- 10 keV X-ray concentrators.  The detection of fading 
X-ray emission, combined with the high sensitivity and the
ability of the concentrators to
refine the position to the arcminute level, led to the subsequent discovery of
long lived emission at lower frequencies \cite{c+97,JvP+97,f+97}.

Optical spectroscopy of the afterglow of GRB 970508 led to
the definitive demonstration of the extragalactic nature of this GRB
\cite{mdk+97}.  The precise positions provided by radio and/or optical
afterglow observations have allowed for the identification of host
galaxies, found in almost every case.  Not only has this provided
further redshift determinations, but it has been useful in tying GRBs
to star formation through measurements of the host star formation rate
(e.g. \cite{kdr+98,dkb+98}). HST with its exquisite resolution has been
critical in localizing GRBs within their host galaxies and thereby shed
light on their progenitors (e.g. \cite{fpt+99,hh99,bod+99}).
Observations of the radio afterglow have directly established the
relativistic nature of the GRB explosions \cite{f+97} and provided
evidence linking GRBs to dusty star-forming regions. Radio observations
are excellent probes of the circumburst medium and the current evidence
suggests that the progenitors are massive stars with copious stellar
winds.  The latest twist is an apparent connection of GRBs with SNe
\cite{bkd+99}. Separately, an important development is the possible
association of a GRB with a nearby (40 Mpc) peculiar SN \cite{gvv+98,kfw+98}.

\begin{figure}[htb] 
\centerline{\psfig{figure=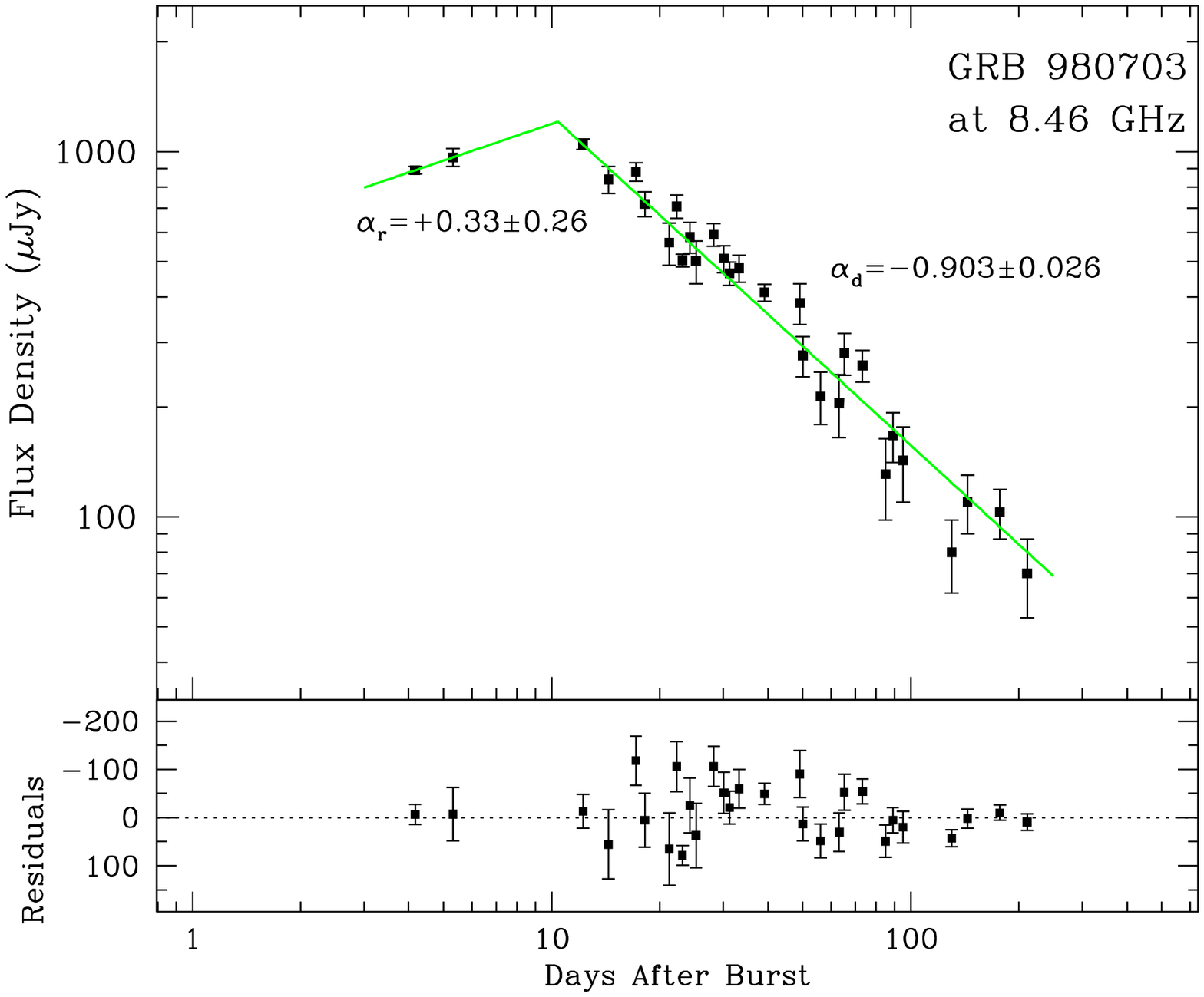,width=7.3cm}\qquad\qquad\psfig{figure=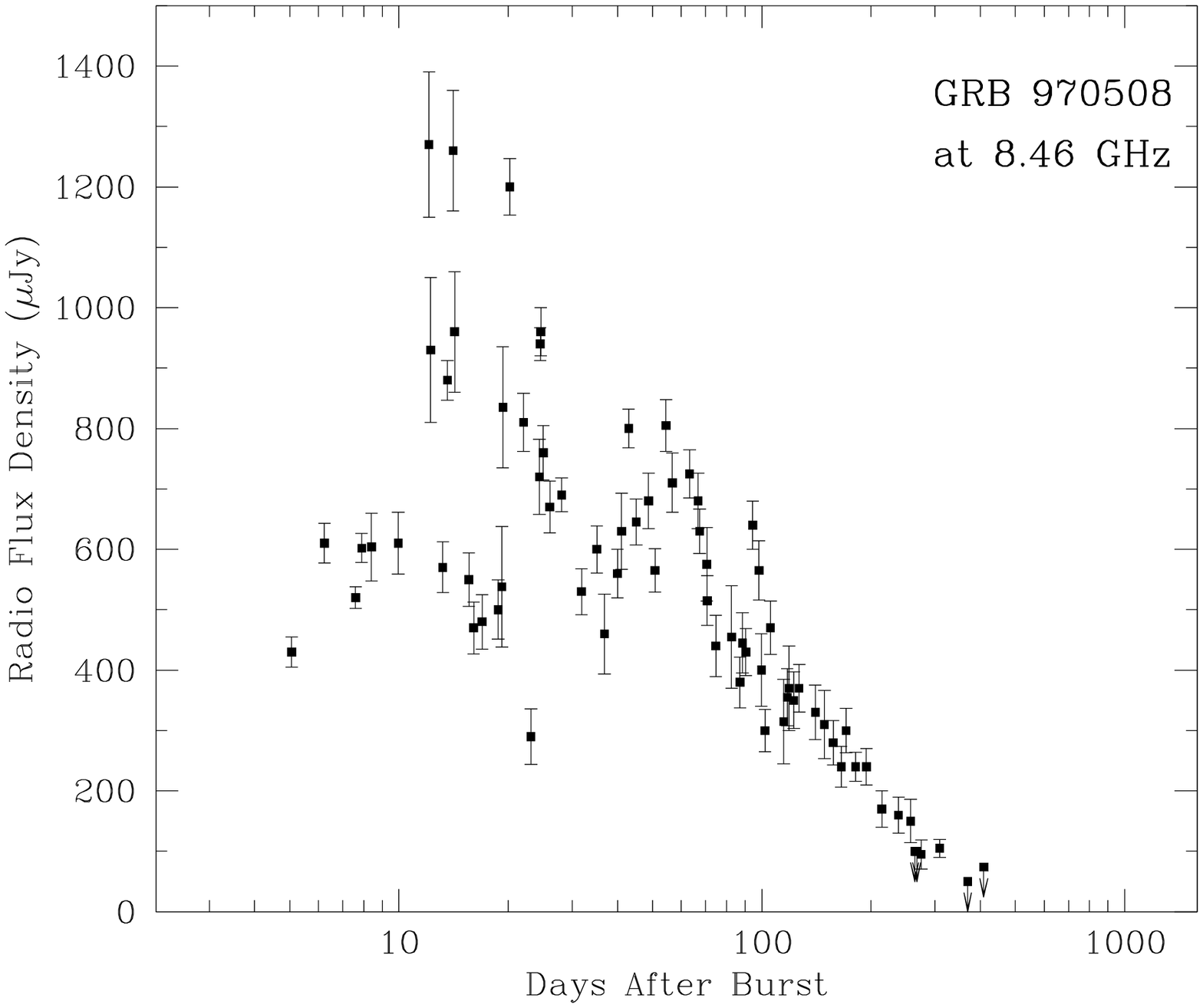,width=7.3cm,angle=0}}
\vspace{10pt}
\caption[]{\small {\it
    Left: The radio light curve of GRB 980703. This is a typical
    afterglow, a rise to a peak followed by a power law decay.  The
    longer lifetime of the radio afterglow allows us to see both the
    rise and the fall of the afterglow emission. In contrast, at
    optical and X-ray emission, most of the times we see only the
    decaying portion of the light curve.  Right: The radio light curve
    of GRB970508 \cite{fwk00}.  The wild fluctuations of the light
    curve in the first three weeks are chromatic.  At later times, the
    fluctuations become broad-band and subdued.  These fluctuations
    are a result of multi-path propagation of the radio waves in the
    Galactic interstellar medium. As the source expands (at
    superluminal speeds) the scintillation changes from diffractive to
    refractive scintillation. This is analogous to why stars twinkle
    but planets do not.\label{fig:980703-970508}}}
\end{figure}

In this paper we review the primary advances resulting from afterglow
studies. \S{II} discusses the statistics of detections to-date,
including possible causes for the lack of radio and optical afterglows
from some GRBs. In \S{III} we review constraints on the nature of the
progenitor population(s), in particular evidence linking some classes
of GRBs to SNe.  \S{IV} describes the status of current understanding
of the physics of the afterglow emission.  Here we compare
observations to predictions of the basic spherically-symmetric model,
and describe complications arising from deviations from spherical
symmetry and non-uniform distribution of the circumburst medium.  We
conclude with speculations of the near and long-term advances in this
field (\S{V}).

We point out that this review has two biases. First, given the
concentration of previous review articles on optical and X-ray
observations, we emphasize the unique contributions of radio afterglow
measurements. Second, this article is intended to also provide a
summary of the efforts of the Caltech-NRAO-CARA GRB collaboration, and
therefore details our work in particular. This review is in response
to review talks given at the 1999 Maryland October meeting (SRK) and
the 5th Huntsville GRB meeting (DAF and SRK).

\section{Statistics of Afterglow Detections}
\label{sec:statistics}

Afterglow emission was first detected from GRB~970228, both at X-ray
\cite{c+97} and optical frequencies \cite{JvP+97}, but not
at radio wavelengths \cite{fksw98}.  The first radio afterglow
detection came following the localization of GRB 970508
\cite{f+97}.  Figure~\ref{fig:980703-970508} shows
two examples of radio lightcurves. The radio afterglow of GRB 970508
is famous for several reasons: it was the first radio detection, it
gave the first direct demonstration of relativistic expansion, and it
remains the longest-lived afterglow \cite{fwk00}.

Afterglow emission is now routinely detected across the
electromagnetic spectrum. BeppoSAX has been
joined in studying the X-ray afterglows 
by the All Sky Monitor (ASM) aboard the X-ray Timing Explorer
(XTE), the Japanese ASCA mission, and recently the Chandra X-ray
observatory (CXO).  A veritable armada of
optical facilities (ranging from 1-m class telescopes to the 10-m Keck
telescopes) routinely discover and study optical afterglows.  
The HST has been primarily used to make exquisite images of 
the host galaxies (see above) but in the near future we expect other uses
such as UV spectroscopy and identification of underlying SNe.
The VLA
has led the detection in radio. However, other
centimeter-wavelength facilities (the Australia Telescope National
Facility, Westerbork Synthesis Radio Telescope, the Ryle Telescope)
and millimeter wavelengths (James Clerk Maxwell Telescope, the Owens
Valley Millimeter Array, IRAM and the Plateau de Bure Interferometer)
are now regularly contributing to afterglow studies.

\begin{figure}[ht!] 
\centerline{\epsfig{file=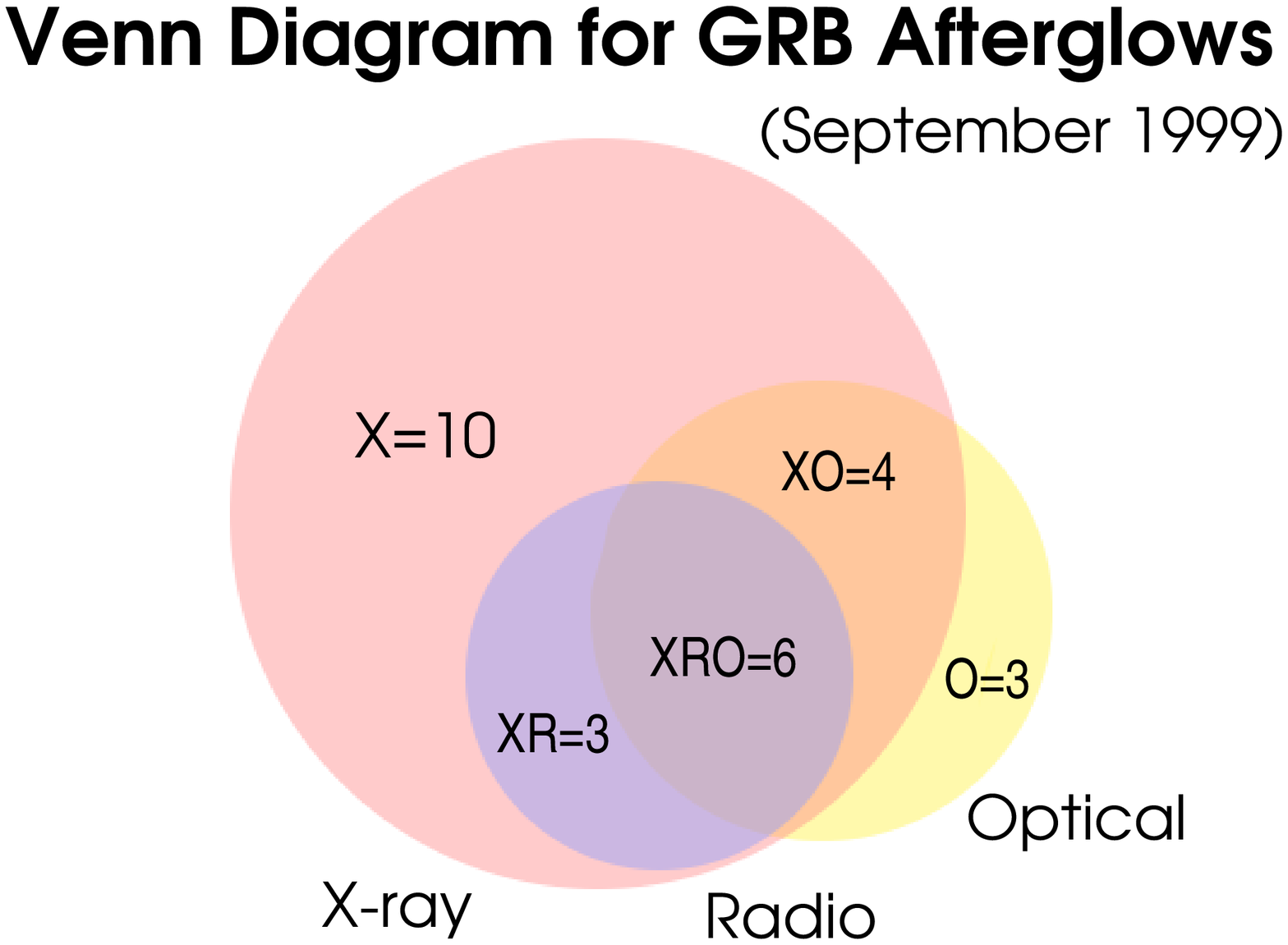,width=3.5in}}
\vspace{10pt}
\caption[]{\small {\it
    A Venn diagram showing the detection statistics for 26
    well-localized GRBs in the Northern and Southern hemispheres. Of
    the 23 GRBs for which X-ray afterglows have been detected to date,
    10 have optical afterglows (XO + XOR) and 9 have radio
    afterglows (XR + XOR). In total there are 13 optical and/or radio
    afterglows with corresponding X-ray afterglows.
\label{fig:venn}}}
\end{figure}

Figure~\ref{fig:venn} summarizes the statistics of afterglow
detections. In almost all cases, X-ray emission has been detected,
establishing the critical importance of prompt X-ray observations.
Optical afterglow appears to be detected in about 2/3 of all
well-localized events if sufficiently deep optical images are taken
rapidly (i.e. within a day or so of the burst).  Radio afterglows are
detected in 40\% of the cases -- far more often than usually assumed.
We refer the reader to the Frail {\it et al.} \cite{fkw+00} for a
comprehensive summary of the X-ray/optical/radio afterglow detection
statistics. The non-detections are, as discussed below,
as interesting as
the detections.

\smallskip
\noindent{\bf Radio Non-detection.}
The failure to find radio afterglow is most likely due to lack of
sensitivity.  The brightest radio afterglow to date is that from GRB
991208 (Frail GCN\footnote{GCN refers to the GRB Coordinates Network
Circular Services. This network is maintained by S.  Barthelmy at the
Goddard Space Flight Center;\\ see {$\rm
http://lheawww.gsfc.nasa.gov/docs/gamcosray/legr/bacodine/gcn\_main.html$}}
451) with a peak flux of 2 mJy, a 60-$\sigma$ detection (at centimeter
wavelengths) whereas the weakest afterglow is typically around
5$\sigma$.  In contrast, at optical and X-ray wavelengths, afterglow
emission is routinely detected at hundreds of sigma. If the VLA were
to be upgraded by a factor of 10 in sensitivity, then we predict
that radio afterglow emission would, like X-ray emission, be detected
from most GRBs.

\smallskip
\noindent{\bf Optical Non-detection.} Non-detection at optical wavelengths
is more interesting, as it may result in some cases from extinction
along
the line of sight or within the source.
Bad weather as well as rapid fading of the afterglow has certainly
hindered some optical searches, which, due to notification delays,
typically begin some hours after the event. Furthermore, low Galactic
latitude events may be obscured, or hidden in crowded foregrounds.
However, in some cases deep searches have been performed with no
success.  Here, non-detection likely results from 
extinction by dust in the burst host galaxy and/or absorption by the
intergalactic medium. GRB 970828 \cite{ggv+98} is one example, as is
the more dramatic case of GRB 980329.  This burst was one of the brightest
events in the WFC\cite{iaa+98}.  Searches for optical afterglow
emission failed to identify any counterpart.  VLA 
observations identified an unusual radio variable in the field
\cite{tfk+98}.  Soon thereafter, a red afterglow and a bright IR
afterglow were identified (Klose GCN 43, Larkin et al. GCN 44). Taylor
et al.  \cite{tfk+98} suggest that the GRB arose in a region with high
extinction.  Further optical and IR work on this interesting afterglow
can be found in
\cite{gcp+99}, \cite{ppm+98}, and \cite{rlm+99}.

Optically dim ``red'' but bright IR afterglows
can also result from the GRB being located at high redshift.  Intergalactic
HI absorption will result in a wavelength cutoff below the Lyman
limit, $<912(1+z)$ \AA, where $z$ is the redshift of the source.  This
effect was originally invoked to explain the faint R-band but bright
IR emission from GRB 980329 \cite{f99}.  We now know, based on recent
Keck observations, that the GRB host is blue, incompatible with a
high-$z$ origin.  Rather, it is more tenable that the host is a
typical star-forming galaxy with dusty star-forming regions, and that
the GRB occurred in one such region\cite{tfk+98}.  We are presently
carrying out IR spectroscopy of this host to determine the redshift
and the star formation rate (SFR).  While searching for ``R dropouts''
may in the future provide an effective method for finding high-redshift
events, it is clear that cross-calibrated multi-band photometry of
higher quality than currently exists will be required to make this
useful.

\smallskip
\noindent{\bf X-ray Non-detection.}
The spectra of most GRB events clearly extend into the X-ray band, as
established by {\it GINGA} observations\cite{sfmy98}.  How the X-ray
emission observed during the burst connects to the X-ray afterglow is
uncertain, due to sensitivity limitations of wide-field monitors.
X-ray afterglow emission appears to be ubiquitous. Observations
of the X-ray afterglow are important for two reasons: (i) the
observations of the X-ray afterglow by sensitive imaging instruments
(e.g. the concentrators aboard BeppoSAX) result in sufficiently
precise (arcminute) localization and (ii) a significant (perhaps even
a dominant) fraction of  the explosion
energy appears to be radiated in this band.  Of all the SAX
bursts, GRB 970111 is peculiar for the absence of X-ray afterglow
(admittedly the data were obtained about 17 hours after the burst)
\cite{fac+99}.  In view of the critical role played by
X-ray afterglow in localization of GRBs we 
regard this non-detection to be worthy of further
investigation.

\begin{table*}
\caption[]{Basic Properties of Selected GRBs}
\begin{tabular}{ccccccl}
GRB & $\alpha$(J2000) &$\delta$(J2000)  & R$_{\rm host}$ & S $\times
10^{-6}$  & $z$ & References\tablenote{References to redshift
determination.} \\
\omit& (h m s) & ($^{\circ}~'~''$) & (mag) & (erg cm$^{-2}$) &\omit & \omit \\
\tableline
970228  & 05 01 47 & +11 46.9 & 25.2\tablenote{$V$-band magnitude from
  HST. All others are R magnitude in the Johnson system.}
 & 1.7 & 0.695   & Djorgovski et al. GCN 289\\
970508  & 06 53 49 & +79 16.3 & 25.7  &
          3.1    & 0.835 & \cite{mdk+97,bdk+98}\\
970828  & 18 08 32 & +59 18 52 & TBD      &
          74     & 0.957 & \cite{dfk+2000}\\
971214  & 11 56 26  & +65 12.0 & 25.6        & 
          11     & 3.418 & \cite{kdr+98}\\
980326  & 08 36 34 & $-$18 51.4  & $\age 27.3$ & 
           1 & \ldots & \omit\\
980329  & 07 02 38 & $+$38 50.7  & 25.4 &  
          50 & \ldots & \omit\\
980519  & 21 22 21 & $+$77 15.7  & 26.2 & 
           25 & \ldots & \omit\\
980613  & 10 17 58 & $+$71 27.4 & 24.5 & 
         1.7 & 1.096 & Djorgovski et al. GCN 189 \\
980703  & 23 59 07 & $+$08 35.1 & 22.6 & 
          37 & 0.966 & \cite{dkb+98}\\ 
981226  & 23 29 37 & $-$23 55 54 & $\age 22$ & 
          N.A. & \ldots & \omit\\
990123  & 15 25 31 & +44 46 00 & 24.4 & 
        265 & 1.600 & \cite{kdo+99} \\
990510  & 13 38 07 & $-$80 29 49 & $\age 28$ &
         23 & 1.619 & Vreeswijk et al. GCN 324\\
990712  & 22 31 53 & $-$73 24 29    &  21.78 & 
        N.A.& 0.430 & Galama et al. GCN 388\\
991208  &16 33 54 & +46 27 21 & $\age 25$ &
         100 & 0.706 & Dodonov et al. GCN 475\\
991216  & 05 09 31 & +11 17 07 & 24.5 & 
        256 & 1.020 & Vreeswijk et al. GCN 496\\
\end{tabular}
\end{table*}
\smallskip

\section{The Nature of the Progenitors}
\label{sec:progenitors}

\begin{figure}[ht!]
\centerline{\epsfig{figure=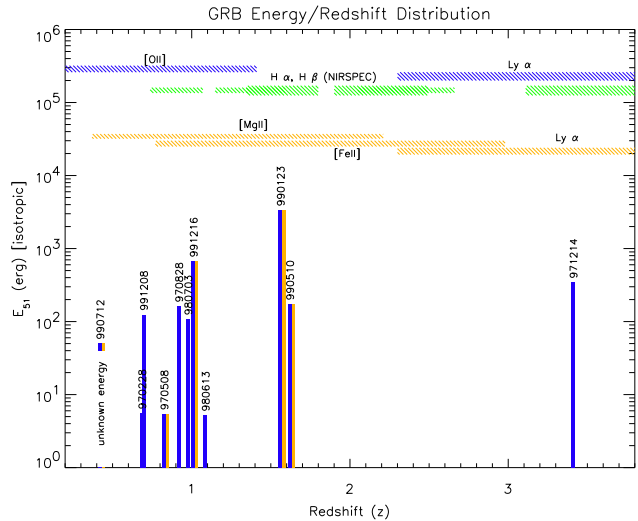,width=9.cm}}
\vspace{7pt}
\caption[]{\small {\it
    The isotropic gamma-ray energy distribution of GRBs with confirmed
    redshifts. Bursts indicated in black are those with
    spectroscopically confirmed emission lines from the host galaxies;
    bursts indicated by a shaded column (e.g. 990123) are those with
    absorption line redshifts. The relevant key absorption or emission
features are noted at the top of the figure.
\label{fig:redshift}}}
\end{figure}

In almost all cases, a host galaxy has been identified at the location
of the fading afterglow. GRB redshifts can be obtained either via
absorption spectroscopy (when the transient is bright) or by emission
spectroscopy of the host galaxy.  In Figure~\ref{fig:redshift} and
Table~1 we summarize the measured redshifts and host galaxy
magnitudes. 
While the distance scale debate is settled (at least for the class of
long duration GRBs, see below) 
we remain relatively ignorant of the nature of the central
engine.  Currently popular GRB models fall into two categories: {(i)}
the coalescence of compact objects (neutron stars, black holes and
white dwarfs
\cite{eic+89,ls74,moc93,npp92}) and {(ii)} the collapse of the central
iron core of a massive star to a spinning black hole, a ``collapsar''
\cite{woo93,mw99}.  We now summarize the light shed on the progenitor
problem by afterglow studies.

\smallskip
\noindent{\bf The Location of GRBs Within Hosts.}
A fundamental insight into the nature of SNe came
from their location with respect to other objects within the host
galaxy (specifically HII regions and spiral arms) and the morphology
of the host galaxy itself (elliptical versus spiral).  
In a similar manner, we are now making progress in understanding GRB
progenitors by measuring offsets with respect to other objects in the
host galaxies.  The rather good coincidence of GRBs with host galaxies
already suggests that they are unlikely to be a halo population (as
would be expected in the coalescence scenario \cite{bsp99}). 
On the other hand,
with the possible exception of GRB~970508\cite{pfb+99}, they are clearly not
associated with galactic nuclei (i.e. massive central black holes).
Typical offsets of GRBs from the centroid of their host galaxies are
comparable to the
half-light radii of field galaxies at comparable magnitudes,
suggesting that GRBs originate from stellar populations.

\smallskip
\noindent{\bf Host Galaxies.}
Demonstrating a direct link between GRBs and (massive) star formation
is more difficult.  On the whole, the population of identified hosts
seems typical in comparison to field galaxies in the same redshift and
magnitude range.  The hosts have average luminosities for field
galaxies, modulo corrections due to evolution.  Their emission line
fluxes and equivalent widths are also statistically
indistinguishable 
from the normal field galaxy population.  The observed star formation
rates, derived from recombination line fluxes (mostly the [O II] 
3727$\,\AA$ 
line) and from the UV continuum flux range from less than $1\,
M_\odot$ yr$^{-1}$ to several tens of $M_\odot$ yr$^{-1}$ --  
typical of normal
galaxies at comparable redshifts (extinction corrections can increase
these numbers by a factor of a few, but similar corrections apply to
the comparison field galaxy population as well).  It will probably be
necessary to have a sample of several tens of GRB hosts before a
correlation of GRBs with the (massive) star formation rate can be
tested statistically.  
However, below we point to several specific
examples which are suggestive of a link between GRBs and star-forming
regions.

\smallskip
\noindent{\bf Association with Starforming Regions.}
There is evidence showing that GRBs arise from 
dusty regions within their host galaxies. In this 
respect, radio observations provide a unique tool for detecting
events in regions of high ambient density (as was the case for GRB
980329).  An even more extreme example is GRB 970828, where the host
was identified based {\it solely} on the VLA discovery of a radio
flare \cite{dfk+2000}. Interestingly enough, this is the dustiest galaxy in
the sample of GRB hosts to-date.

Second, some GRBs appear to be located within identifiable
star-forming regions.  An example is GRB~990123 \cite{bod+99,ftm+99,hh99}.
VLA observations of GRB 980703 \cite{fbk+2000} are perhaps more
convincing. The radio observations can be sensibly interpreted by
appealing to free-free absorption from a foreground HII region (which
would dwarf the Orion complex).  If this interpretation is correct
then this would be strong evidence for a GRB being located within a
starburst region.

\begin{figure}[ht!]
\centerline{\psfig{figure=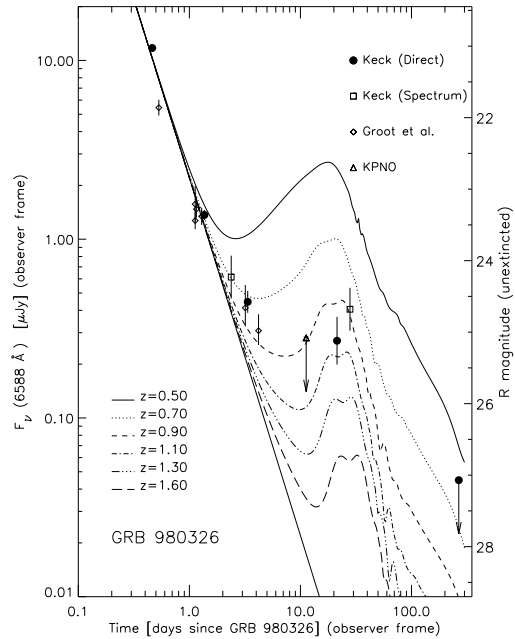,width=7.0cm}}
\vspace{7pt}
\caption[]{\small {\it R-band light curve of GRB\,980326 and the
    sum of an initial power-law decay plus Ic supernova light
    curve for redshifts ranging from $z = 0.50$ to
    $z=1.60$; from \cite{bkd+99}.  
\label{fig:SN-curves}} \vspace{0.1cm}}
\end{figure}

\medskip
\noindent{\bf The GRB--SN link.}
If GRBs arise from the collapse of a massive star, it is an
unavoidable consequence that emission from the underlying supernova
should be superimposed on the afterglow.  Bloom et al. \cite{bkd+99}
may have made the first detection of a possible SN component
in the GRB~980326
lightcurve (Fig. \ref{fig:SN-curves}).  These authors noted that SNe,
in contrast to afterglows, have distinctive temporal and spectral
signatures: rising to a maximum at $\sim 20(1+z)$ days, with little
emission blueward of about 4000\,\AA\ in the restframe
(and certainly blueward of
3000\,\AA) owing to a multitude of resonance absorption lines.  This
discovery has led to other possible SN detections, most notably
GRB\,970228 \cite{gal+99,rei99}.

The suggestion of a GRB--SN connection is an intriguing one but it has
yet to be placed on a firm footing.  Important questions are: (i) are
all long-duration GRBs accompanied by SNe?  (ii) if so, are these SNe
of type Ib/c?  Ground-based observations are possible in those cases
where the afterglow decays rapidly (e.g. GRB 980326) or if high quality
optical and IR observations exist (e.g. GRB 970228).

We need more examples to test the GRB--SN link.
Future progress will depend on a combination of ground and HST
observations.  For relatively nearby GRBs especially those with 
a rapidly decaying optical afterglow
it would be attractive and
feasible to obtain the spectrum of the SN around the time when the
flux from the SN peaks.
A moderate quality
spectrum with SN-like features would have the singular
advantage of definitively
confirming the SN interpretation (as opposed to alternatives involving
re-radiation by dust \cite{wd99}).  However, for most GRBs, we expect
HST observations to play a critical role. 
HST's widely recognized strengths in accurate photometry of
sources embedded in galaxies\cite{gkc+98} and photometric stability
make the detection of a faint SN against the optical afterglow and the
host galaxy possible.

\medskip
\noindent{\bf Diversity of the Progenitor Population.}
As was the case with SNe, it is likely naive to think of a single
progenitor population.  
Below, we discuss the
two additional classes which show some promise:
the mysterious short duration GRBs
and a possible class of low luminosity GRBs associated with SNe.

\begin{figure}[ht!]
\centerline{\epsfig{figure=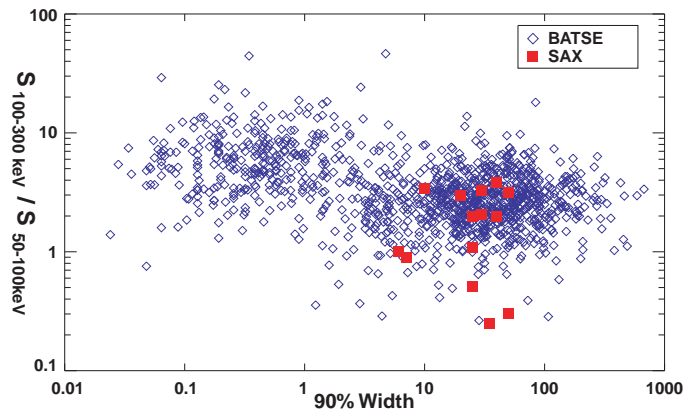,width=9.cm}}
\vspace{7pt}
\caption[]{\small {\it
    Distribution of duration ($T_{90}$) vs. spectral hardness for
    BATSE bursts (diamonds) from the 4B catalogue. There is a clear
    suggestion of two groups of GRBs: short/hard and long/soft
    events. Events localized by BeppoSAX
    (solid squares) appear to belong to the long duration
    class.\label{fig:bim}}}
\end{figure}

\smallskip
\noindent
{\em Short Events.}  It has been known for some time that the
distribution of the duration of GRBs appears to be bimodal
\cite{fm95}; see Figure~\ref{fig:bim}.  Furthermore, these two groups
may have different spatial distributions \cite{kc96b}, with the short
bursts being detected out to smaller limiting redshifts.  However, we
know very little about this class of GRBs since, as noted earlier, all
bursts localized by BeppoSAX and RXTE thus far are of long duration
(Figure \ref{fig:bim}).  Fortunately, improvements in BeppoSAX and the
imminent launch of HETE-2 provide for the first time the opportunity to
follow-up short GRBs.

The short duration bursts are difficult to accommodate in the
collapsar model, given the long collapse time of the core.  However,
they find a natural explanation in the coalescence models. How would
these bursts manifest themselves?  Li \&\ Paczy\`nski
\cite{lp98b} speculate that if the short-duration bursts result
from NS--NS mergers then they may leave a bright, but short-lived
($\ale 1$ day) optical transient.  Radio observations provide a
complementary tool for determining the nature of the short duration
bursts.  The low ambient density would result in weak afterglows (since
flux $\propto \rho^{1/2}$) which are potentially detectable.  Radio
observations have additional advantages of a longer lived afterglow,
immunity from weather and freedom from the diurnal cycle.

\begin{figure}[t!]
\centerline{\psfig{figure=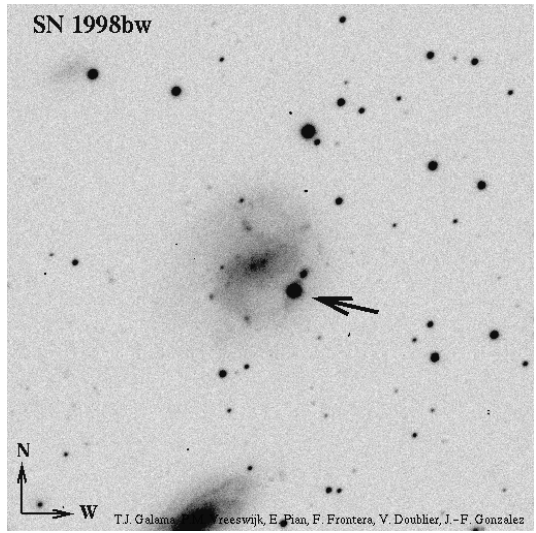,width=7.0cm,angle=0}}
\vspace{7pt}
\caption[]{\small {\it Discovery image of SN 1998bw\cite{gvv+98}.
The SN is the bright object (marked with an arrow) SW of the
nucleus. Relative to typical SNe, this SN is more energetic and 
appears to have synthesized ten times more Nickel.
\label{fig:sn1998bw}}
\vspace{0.1cm}}
\end{figure}

\begin{figure}[ht!]
\centerline{\epsfig{figure=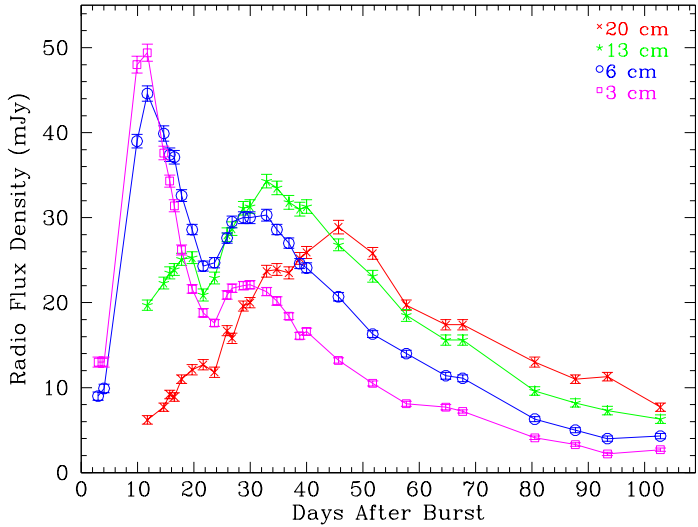,width=8.cm,angle=-90}}
\caption[] {\small {\it
    The radio light curve of SN 1998bw at four wavelengths
    \cite{kfw+98}.  The peak brightness temperature of SN 1998bw at
    early times is $10^{13}$ K, well in excess of the inverse Compton
    limit of $5\times{10}^{11}$ K, and can be best understood if the
    radio emission originates from a relativistic shock ($\Gamma\age 2$).
\label{fig:sn1998bw-lc}}}
\end{figure}

\smallskip 
\noindent{\em Gamma-ray Bursts Associated with Supernovae.} Observers
and theorists alike have been intrigued by the possibility that the
bright supernova, SN~1998bw, discovered by Galama et al.
\cite{gvv+98} in the error circle of GRB 980425 \cite{paa+99}, is
associated with the gamma-ray event (Figure~\ref{fig:sn1998bw}). 
Kulkarni et al. \cite{kfw+98}
discovered that the SN had an extremely bright radio counterpart; see
Figure \ref{fig:sn1998bw-lc}. We noted that the inferred brightness
temperature exceeded the inverse Compton catastrophe limit of $5\times
10^{11}$ K and to avoid rapid cooling we postulated the existence of a
relativistically expanding blastwave ($\Gamma\age 2$). This
relativistic shock is, of course, in addition to the usual
sub-relativistic SN shock.  This relativistic shock may have produced
the GRB at early times. (We note here that we disagree with the much
lower energy estimates of \cite{wl99}; our recent calculations using
the same assumptions as those made in \cite{wl99} result in an energy
estimate similar to that obtained earlier \cite{kfw+98} from
minimum-energy formulation).
The optical modeling of the
lightcurve and the spectra show that GRB 980425 was especially
energetic \cite{imn+98,wes99} with an energy release of $3\times 10^{52}$
erg and Nickel production of nearly nearly a solar mass.

If GRB~980425 is associated with 1998bw, then this type of event is
rare among the SAX localizations.  GRB 980425 is most certainly not a
typical GRB: the red-shift of SN 1998bw is 0.0085 and the $\gamma$-ray
energy release in GRB 980425 is at least four orders of magnitude less
than in other cosmologically located GRBs.  For this reason, most
astronomers (especially those in the GRB field; see Wheeler's foray in
experimental sociology\cite{w99}) do not believe the association
between GRB 980425 and SN 1998bw. On the other hand, as evidenced by 
the 
intense interest in and modeling of the radio and optical data of SN
1998bw, this object is of considerable interest to the SN
community.  Indeed, we 
believe that the proposed GRB--SN association controversy has
muddied the main issue: SN 1998bw is an interesting SN in its own
right.

What is the true distinguishing feature of SN 1998bw that may connect
it to a GRB event? Is it the large energy release, as suggested by
several authors\cite{imn+98,grs+99}?  We argue that in fact it is the
energy {\em coupled into relativistic ejecta} that most closely
connects SN 1998bw to a GRB. In a typical SN, about $10^{51}$ erg is
coupled to the envelope of the star (a small fraction of the total SN
energy release of $10^{53}$ erg).  In a GRB, a similar amount of
energy ($10^{51}$--$10^{52}$~erg depending on the event)
is coupled to a much smaller ejecta mass, resulting in relativistic
outflow.  For SN 1998bw, applying the minimum energy formulation
to the radio observations we infer the relativistic shell
to contain $\sim 10^{50}$ erg. 
Not only is this uncharacteristic
of a typical SN (there exists no evidence for relativistic ejecta in ordinary
SN), but it is not dissimilar from the energy implied for GRB
outflows.  One could therefore envisage a continuum of physical
phenomenon between SN 1998bw and cosmological GRBs provided we use the
energy in the relativistic ejecta as the basic underlying
parameter and not the isotropic gamma-ray release.

\section{Afterglow: The Physics and Energetics of the Fireball}
\label{sec:afterglow}

One can consider a GRB to be like a SN explosion with a central source
releasing energy $E_0$ (comparable to the mechanical release of energy
in an SN). This is the so-called fireball model.
The difference between an SN and a GRB is primarily in
ejecta mass: 1--10 $M_\odot$ for SNe whereas only $10^{-5}\,M_\odot$
for GRBs. The evolution of a GRB is much faster than that of a SN due
to two factors: the ejecta expand relativistically and, thanks to the
smaller ejecta mass, the optical depth is considerably smaller.

As the ejecta encounter ambient gas, two shocks are produced: a
short-lived reverse shock (traveling through the ejecta) and a
long-lived forward shock (propagating into the swept-up ambient gas).
Afterglow emission is identified with emission from the forward shock.
In order to obtain significant afterglow emission, several conditions
are necessary. (1) Rapid equipartition of electrons with the shocked
protons (which hold most of the energy). (2) Acceleration of electrons
to a power law spectrum (particle Lorentz factor distribution,
$dN/d\gamma\propto \gamma^{-p}$). (3) Rapid growth of the magnetic field
with energy density in the range of $10^{-2}$ of that of the protons.
Under these circumstances, afterglow emission is dominated by
synchrotron emission of the accelerated particles; see
\cite{spn97,w97a}.  The weakness of this model is the assumption of
growth in the magnetic field strength to the high values noted above
(R. Blandford, pers. comm.).

\begin{figure}[!ht]
\centerline{\epsfig{figure=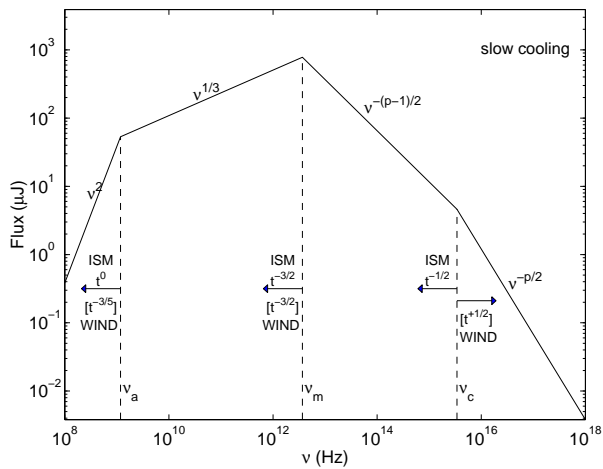,width=8cm}}
\vspace{10pt}
\caption[]{\small{\it
    Broad-band spectrum ($f_\nu$) of the afterglow from a spherical fireball
    with constant density (``ISM'' model; see text) and $\rho\propto
    r^{-2}$ medium (``wind'' model; see text). This is representative of the
    observed spectrum few days after the burst.  Note the distinct
    evolution of $\nu_{a}$ and $\nu_c$ in the two
    models.\label{fig:sari}}}
\end{figure}

The theoretically expected afterglow spectrum is shown in
Figure~\ref{fig:sari}. Three key frequencies can be identified:
$\nu_a$, the synchrotron self-absorption
frequency; $\nu_m$, the frequency of the electron with a minimum
Lorentz factor (corresponding to the thermal energy behind the shock)
and $\nu_c$, the cooling frequency. Electrons which radiate above
$\nu_c$ cool on timescales equal to the age of the shock.  The
evolution of these three frequencies is determined by the
hydrodynamical evolution of the shock which in turn is affected by two
principal factors: the environment of the GRB and the geometry of the
explosion.

\smallskip\noindent
{\em The GRB environment.}  The earliest afterglow models made the
simplifying assumption of expansion into a constant density
medium. This is an appropriate assumption should the GRB progenitor
explode into a typical location of the host galaxy. However, there is
increasing evidence tying GRBs to massive stars (see
\S\ref{sec:progenitors}).  It is well known that massive stars lose
matter throughout their lifetime and thus one expects the circumburst
medium to exhibit a density profile, $\rho\propto r^{-2}$ where $r$ is
the distance from the progenitor.  Chevalier \&\ Li
\cite{cl99} refer to these two models as the ISM (interstellar medium)
and the wind model respectively. As can be seen from
Figure~\ref{fig:sari} these two models give rise to rather different
evolution of the three critical frequencies.

\smallskip\noindent {\em Geometry: Jets versus Spheres.} The
hydrodynamics is also affected by the geometry of the explosion.  Many
powerful astrophysical sources have jet-like structure. There is
evidence (from polarization observations) indicating asymmetric
expansion in SNe \cite{w99},  so it is only reasonable to assume that
GRB afterglows also have jet-like geometry as well. 
A clear determination of the
geometry is essential in order to infer the true energy of the
explosion. This is especially important for energetic bursts such as
GRB 990123 whose isotropic energy release approaches $M_\odot c^2$.

Let the opening angle of the jet be $\theta_0$. As long as the bulk
Lorentz factor, $\Gamma$, is larger than $\theta_0^{-1}$, the evolution
of the jet is exactly the same as that of a sphere (for an observer
situated on the jet axis). However, once $\Gamma$ falls below
$\theta_0^{-1}$ then two effects become important. First, for a well
defined jet, the on-axis observer sees an edge and thus one expects to
see a break in the afterglow emission.  Second, the lateral expansion
of the jet (due to heated and shocked particles) will start affecting
the hydrodynamical explosion.

\smallskip\noindent {\bf Wind or ISM?} 
The two key diagnostics to distinguish these two models are the
evolution of the cooling frequency (see Figure~\ref{fig:sari}) and the
early behavior of the radio emission.   In the wind model, the radio
emission rises rapidly (relative to the ISM model) and the synchrotron
self-absorption frequency falls rapidly with time.  
Both these result
from the fact that the ambient density decreases with radius (and hence
in time)
in the wind model.

\begin{figure}[hb!]
\centerline{\epsfig{figure=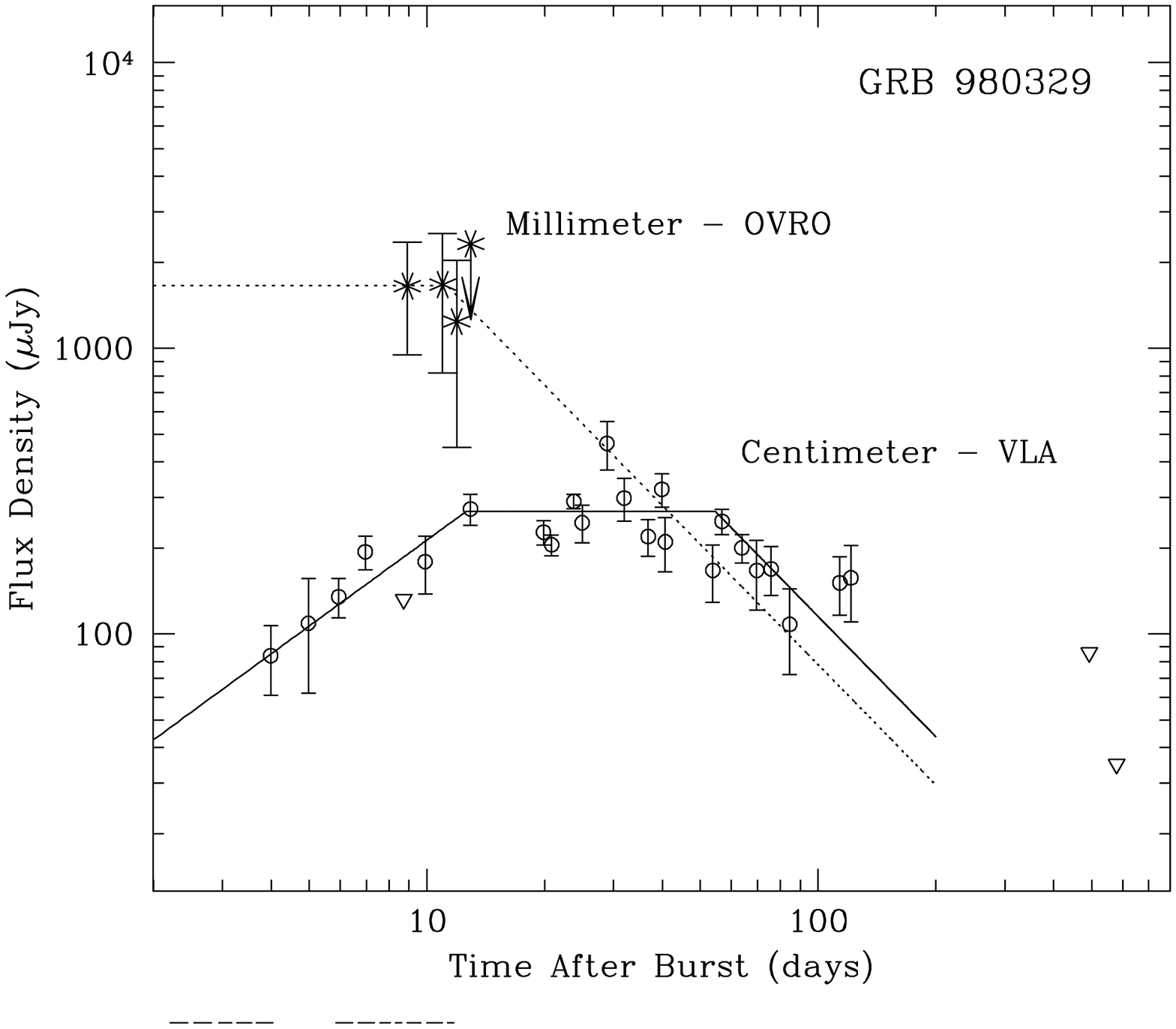,width=7.cm}\qquad\qquad
\epsfig{figure=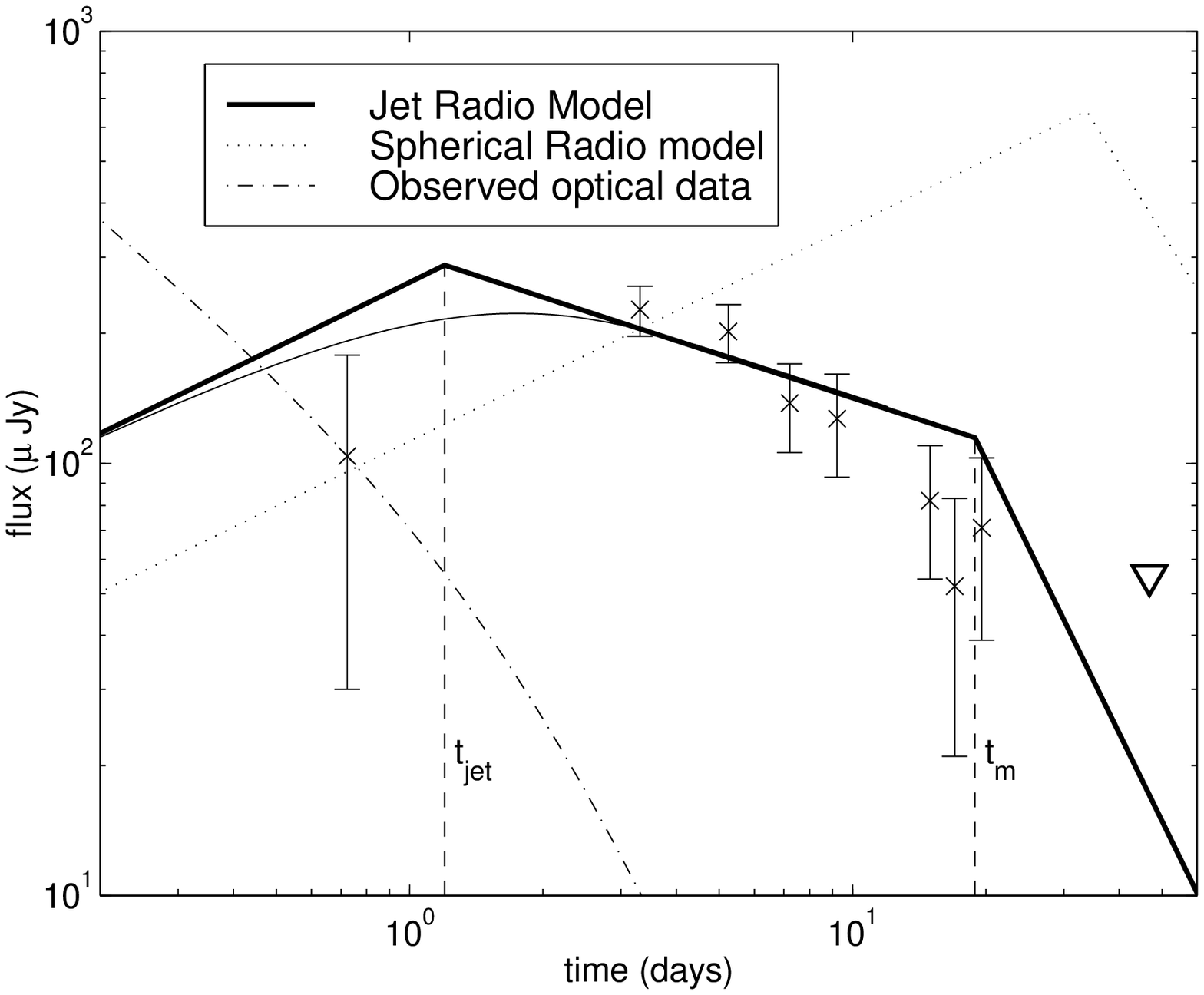,width=7cm,angle=0}}
\vspace{10pt}
\caption[]{\small {\it Left: 
    Radio afterglow of GRB 980329 \cite{fks+99}. 
The rapid rise of the centimeter
    flux and the high absorption frequency (signified by the
    considerable strength of the millimeter emission) offer
    good support for GRB 980329 expanding into a circumburst medium
    with density falling as inverse square distance. 
    The lines represents a wind model based on 
    X-ray, optical, IR, mm and cm data.
    Right:
    Observed and model radio light curves of GRB 990510 \cite{hbf+99}.
	The model predictions for the radio afterglow emission are
	displayed by the solid line (jet fireball model) and dotted
	line (spherical fireball model). The observed optical afterglow
	emission is displayed by the dotted-dashed line; see text for
	more details.
\label{fig:980329-990510}}} 
\end{figure}

Unfortunately, in general, the current data are not of sufficient
quality to firmly distinguish the two models.  For example in GRB
980519, the same optical and X-ray data appear to be adequately
explained by the jet+ISM model \cite{sph99} and the sphere+wind
model \cite{cl99}. Including the radio data tips the balance, but only
slightly, in favor of the wind model \cite{fks+2000}.
In our opinion, the best example for the
wind model is that of GRB 980329 \cite{fks+99}; see
Figure~\ref{fig:980329-990510}.  This afterglow exhibits the two
unique signatures of the wind model: high $\nu_a$ and a rapid rise.
Given the importance of making the distinction between the wind and
the ISM model we urge early wide band radio observations (especially
at high frequencies).

\smallskip\noindent {\bf Energetics.}  Of all the physical
parameters of the fireball, the most eagerly sought parameter is the
total energy $E_0$. By analogy with supernovae, it is $E_0$ which sets
the GRB phenomenon apart from other astrophysical phenomena.
Classes of GRBs may eventually be distinguished and ranked by their
energy budget; for example, long-duration events, short duration
events and supernova-GRBs (see \S\ref{sec:progenitors}).  

One approach has been to use the isotropic $\gamma$-ray energy as a
measure of $E_0$; see Figure~\ref{fig:redshift}.  There are three well
known problems with such estimates. First, collimation of the ejecta
(jets) will result in overestimation of the total energy release.  For
GRB 990510 where a good case for a jet has been established
(Figure~\ref{fig:980329-990510}), the standard isotropic energy
estimate is probably a factor of 300 more than the true energy
\cite{hbf+99}.  Second, even after accounting for a possible jet
geometry, the efficiency of converting the shock energy into gamma-ray
emission is very uncertain. For example, some authors \cite{k99}
advocate low efficiency ($\sim 1\%$) which would result in an enormous
upward correction to the usual isotropic estimates. Third, the bulk
Lorentz factor is extremely high during the emission of $\gamma$-rays
and thus the estimates critically depend on assumption of the geometry
and granularity \cite{kp99} of the emitting region. In particular, if
the emission is from small blobs \cite{kp99} then the inferred
estimates are grossly in error.

In contrast to this highly uncertain situation, afterglows offer (in
principle) more robust methods to evaluate $E_0$. In view of the
importance of determining $E_0$ we summarize the different methods of
determining $E_0$ from afterglow observations.  One approach is to fit
a ``snapshot'' broad-band afterglow spectrum (from radio to X-rays) to
an afterglow model; this approach was pioneered by Wijers \&\ Galama
\cite{wg99}.  The strength of this method is that the estimated $E_0$
is, in principle, robust. Specifically, the estimate does not depend on
the usually unknown environmental factors (run of density).  However,
in practice, this method is very sensitive to the values of the
critical frequencies (Figure~\ref{fig:sari}) which are usually not well
determined. This difficulty explains the wildly differing estimates of $E_0$
for GRB 970508 \cite{wg99,gps99}.  Furthermore, this method uses
measurements obtained at early times (when the afterglow at high
frequencies is bright) with the result that the true source geometry is
hidden by relativistic beaming.

A second approach is to model the light curves of the afterglow in a
given band, specifically a radio band.  The advantages of this method
are the photometric stability of radio interferometers and the low
Lorentz factor at the epoch of the peak of the radio emission. The
disadvantages are two-fold: the sensitivity to the environmental
parameters (density) and the assumption of the constancy of the
microphysics parameters (electron and magnetic field 
equipartition factors).  Application of this
approach to GRB 980703 has resulted in seemingly accurate measures of
the fireball parameters \cite{fbk+2000}.

Freedman \&\ Waxman \cite{fw99} take yet another approach, and
estimate the energy release from late time X-ray observations. They
show that the X-ray flux is insensitive to the GRB environment, and
obtain robust estimates of the fireball energy per unit solid angle:
from $3\times 10^{51}$ erg to $3\times 10^{53}$ erg.

With all the above approaches, however, the possible collimation of the
ejecta in jets is still a major uncertainty.  This can be addressed by
observing the evolution of the afterglow as the ``edge'' of the jet
becomes visible.  In most cases no evidence for jets has been seen,
with the notable exceptions of GRB 990510 and possibly GRB 990123.  In
addition, a variety of statistical arguments (the absence of copious
numbers of ``orphan afterglows'')\cite{gri99,gvb+99,rho97} suggests
that, on average, the collimation cannot be extreme, and that for most
bursts the opening angle is not less than 0.1 radian.  Thus the total
energy for most bursts may be reduced to the range of $10^{50}$ erg to
$3\times 10^{51}$ erg, but could easily be much higher in at least some
cases.

Possibly the best approach to determining the energetics, which
minimizes uncertainties due both to collimation (jets) and to the
environment is to model the afterglow after it becomes
non-relativistic. This method builds on the well established minimum
energy formulation and the self-similarity of the Sedov solution.  Not
only are the ejecta truly non-relativistic, but they are also
essentially spherical, as by this time jets will have had sufficient
time to have undergone significant lateral expansion.
Indeed, we can justifiably call this ``fireball
calorimetry'' \cite{fwk00}.  Applying this technique to the long-lived
afterglow of GRB 970508 (Figure~\ref{fig:980703-970508}) led to the
surprising result that $E_0 \sim 5\times 10^{50}$ erg -- weaker than a
standard SN!  This is an astonishing result. If true, this result would
suggest that it is not $E_0$ which is the prime distinction between
GRBs and SNe but the ejecta mass.  However, Chevalier \&\ Li
\cite{cl2000} interpret the same data in the wind framework and derive
much larger $E_0$.  Clearly, we need more well studied afterglows with
sufficient observations to first distinguish the circumburst
environment (wind versus ISM) and then radio observations over a
sufficiently long baseline to undertake calorimetry. Nonetheless, one
should bear in mind that the current evidence for large energy release
in GRBs is not as strong as is usually assumed.

\section{Epilogue and Future}
\label{sec:epilogue}

Clearly, the GRB field is evolving rapidly. Along what direction[s] will
this field proceed in the coming years?  One way to anticipate the
future is by considering analogies from the past.  

In \S\ref{sec:progenitors} we already discussed the parallels
between the SN field and the GRB field.  Here we discuss the
numerous parallels with quasar astronomy.
First
discovered at radio wavelengths, we now study quasars across the
electromagnetic spectrum.  Although still identified by their
gamma-ray properties, we now recognize the tremendous value of
pan-chromatic GRB and afterglow studies. In both cases, there was
considerable controversy about the distance scale. However, once this
issue was settled, it became clear that quasars are the most energetic
objects (sustained power) whereas GRBs are the most brilliant. For
both, the ultimate energy appears to be related to black holes (albeit
of different masses).

The raging issues in GRB astronomy today are the same that fueled
quasar studies in the 60's: the spatial distribution, the extraction of
energy from the central engine, the transfer of energy from stellar
scales to parsec scales, and the geometry of the relativistic outflow
(sphere or jet).  Astronomers took decades to unify the seemingly
diverse types of quasars, and to conclude that there are two types of
central engines: radio loud and radio quiet.  Likewise, there may well
be two types of GRB engines: rapidly and slowly spinning black holes
emerging respectively from collapse of a rotating core of a massive
star or coalescence of compact objects and the collapse of a massive
star. This picture could potentially explain both the cosmologically
located GRBs and SN 1998bw.  Finally, we can project that in the
future, GRBs may be used to probe distant galaxies, just as quasars are
used today to study the IGM.

There is a feeling in the astronomical community (outside the GRB
community) that the GRB problem is ``solved''. The truth is that the
GRB problem is now getting defined!  We now summarize our view of the
major issues and anticipated near term advances.  In our opinion the
major issues are Diversity, Progenitors and  Energy Generation.

As discussed earlier, high energy observations suggest the
existence of two classes: short and long duration bursts. It is possible
that afterglow observations may demarcate additional classes. If
so, one can contemplate that within a year (assuming abundant
localizations by HETE-2) that we will have new GRB designations such
as {\it s}GRBs (GRBs with late time bump indicative of 
an underlying SN), {\it w}GRBs (GRBs whose
afterglow clearly indicates a wind circumburst medium shaped by stellar
winds), {\it i}GRBs (GRBs which explode in the interstellar medium) and so
on. 

The broad indications are that GRBs are associated with stars and most
likely massive stars. However, we know little beyond this.  Comparing
the unbeamed GRB event rate of $1.8\times 10^{-10}$ yr$^{-1}$
Mpc$^{-3}$ \cite{schmidt99} with $3\times 10^{-5}$ Type Ibc SN
yr$^{-1}$ Mpc$^{-3}$ and $10^{-6}$ yr$^{-1}$ NS--NS merger Mpc$^{-3}$
\cite{lamb99} shows that GRBs events are extremely rare; here we note
that the present data do not support a collimation correction in excess of
100.  It will be quite some time before we will be in a position to
identify the conditions necessary for a star to die as a GRB.

It is our opinion that SN 1998bw is a major development in the field of
stellar collapse. The association (or lack) with GRB 980425
unfortunately has distracted our attention of this important
development. The existence of a significant amount of mildly relativistic
material, $\sim 10^{50}$ erg \cite{kfw+98}, is fascinating and it is
ironic that none of the models can account for this inferred value
whereas most of the theoretical effort has gone into explaining the 
gamma-ray burst itself (especially considering the uncertain association
of GRB 980425 with SN 1998bw). Clearly, SN
1998bw is a rare event but we are convinced that more such events
will be found and
accordingly have mounted a major campaign to identify these SNe. The
robust signatures of this class are high $T_B$ and prompt X-ray
emission since these are necessary consequences of a relativistic
ejecta. We note that if these future events are as bright as SN 1998bw
then the energy in the relativistic ejecta can be directly measured  by
VLBI observations of the expanding radio shell.

It is vitally important to make quantitative progress in determining
the energy release in GRBs. As discussed in \S\ref{sec:afterglow}, firm
estimates of the energy release require well sampled broad-band data
at early times and densely sampled radio light curves out to late
times. This will require a {\it coordinated} approach and necessarily
involve many observatories around the world and in space.  The same
datasets will also help us understand a profound puzzle: if GRBs indeed
arise from the death of massive stars then why do we not see signatures
for a circumburst medium shaped by stellar winds in {\it all} long
duration GRBs? Even ardent supporters of the wind model
\cite{cl99,cl2000} concede that some GRBs (e.g. GRB 990123, 990510) are
due to a jet expanding into a constant density medium.

We now discuss the anticipated  returns. True to our tradition as
observers, we order the discussion by wavelength regimes!

\smallskip
\noindent{\em Radio Observations: Dusty galaxies, Circumstellar Edges
and Reverse Shocks.} Perhaps the most exciting use of radio afterglow
is in identifying dusty star-forming host galaxies. Such host galaxies
are not readily seen at optical wavelengths. Currently, such galaxies
are eagerly sought and studied at sub-millimeter wavelengths. However,
the sensitivity and localization of such galaxies by sub-millimeter
telescopes is poor. In contrast, GRB host galaxies are identified at
the sub-arcsecond level.  The present radio afterglow detection rate of
40\%  already places an upper limit on the amount of star-formation in
dusty regions, viz.  this rate is not larger than that measured from
optical observations.  This result is entirely independent of the
conclusion based on studies in the sub-millimeter regime, or the 
diffuse cosmic FIR background found in the COBE
data.  However, the result does rely on two assumptions: (i) GRBs trace
star formation and (ii) the GRB explosion and  its aftermath does not
radically alter the ambient medium (i.e., with a prompt and complete
destruction of dust grains along the line of sight).

Radio observations of SNe offer a probe of the distribution of the
circumstellar matter. A spectacular example is SN 1980K whose radio
flux dropped 14 yrs after the explosion \cite{mvw+98}.  A progenitor
star which suffered mass loss with variation in the wind speed could
explain the observations.  Indeed, one {\it expects} significant radial
structure in the circumburst medium as the progenitor evolves from a
blue star to a red supergiant and thence to possibly a blue supergiant
etc.  If GRBs come from binary stars which undergo a phase
of common envelope envolution
\cite{bllb99} then the structure would be even more complicated.
Thus radio observations have the potential (in fortunate circumstances)
to give us insight into the mass loss history of the progenitor star[s].

The prompt optical emission from GRB 990123 \cite{abb+99} has been
interpreted to arise from the reverse shock \cite{sp99}. Far less
discussed is the prompt radio emission -- a radio flare -- also seen
from this burst \cite{kfs+99}.  Sari \&\ Piran \cite{sp99}
suggest that the radio emission also originates from the reverse
shock as the electrons cool.  Observations related to the reverse
shocks are important since it is only through these observations that
we have a chance of studying the elusive ejecta.  We now have four
such examples of radio flares \cite{kf00} and this represents an order
of magnitude better success rate than ROTSE+LOTIS. We urge theorists
to pay attention to these new findings. 
More to the point,
radio observations appear to be fruitful for the study of reverse
shocks, especially when combined with observations of the prompt
optical emission. This bodes well for the coming years given the
efforts underway to increase the sensitivity of ROTSE \cite{abb+99}.

\smallskip
\noindent{\em X-ray Observations: Diversity \&\ Progenitors.} 
{\it GINGA} identified a number of X-ray rich GRBs. BeppoSAX has found
several such examples with some bursts lacking significant gamma-ray
emission -- the so-called X-ray flashes \cite{heise99}.
We know very little about these X-ray transients. 
Could they be GRBs
in a very dense environment (with red giant progenitors)?  We need to take
such transients more seriously and intensively followup on such
bursts.

Another interesting finding from {\it GINGA} was the discovery of
precursor soft X-ray emission \cite{m+91}. There is no simple
explanation for this phenomenon in the current internal-external shock
model. We suggest that the soft X-ray emission precursor is similar to
the UV breakout of ordinary SNe. This hypothesis can be confirmed or
rejected by obtaining the redshift to such bursts.

The X-ray rich GRB 981226 \cite{faa+2000,fkb+99} was marked with two additional
peculiarities: a precursor emission and afterglow emission which is
seemingly undetectable after about 12 hours but then rises rapidly
before commencing decay.  Above we alluded to the fact that massive
stars do not have a single phase of mass loss but instead
have a veritable history of
mass loss (from birth to death). The X-ray observations of GRB 981226
could be accounted for in a model in which the progenitor has first a red
supergiant wind followed by a blue supergiant wind.

\smallskip
\noindent{\em Optical Observations: SN link, Short bursts \&\ Geometry.}
The GRB--SN connection is best probed by optical observations.  The
value of optical observations has already been demonstrated by the
current observations of GRB 980326 and 970228. Clearly, more
observations are needed to establish this link. Once this link is
established then one can undertake detailed spectroscopic studies of the
SN with
large ground-based telescopes and photometric studies with HST.

Offsets of GRBs and the morphology of the host galaxies will continue
to be of great interest.  Such observations will help us differentiate
whether some GRBs come from nuclear regions or always from
star-forming regions.  Under the current paradigm, the discovery of
GRBs coincident with elliptical galaxies would be a major surprise.
On the other hand, one expects short bursts to arise in the halo of their
galaxies and thus in this case no coincidence is expected. We expect
HETE-2 to contribute significantly to these issues.
Finally, polarization measurements offer a very convenient way to probe the
geometry of the emitting region as has already been demonstrated
from the discovery of polarization in GRB 990510 (e.g. \cite{lcg99,wvg+99}).

\smallskip
\noindent{\em Acknowledgments.} {\small Our research is supported by
NASA and NSF. JSB holds a  Fannie \&\ John Hertz Foundation Fellowship,
AD holds a Millikan Postdoctoral Fellowship in Experimental Physics,
TJG holds a Fairchild Foundation Postdoctoral Fellowship in Observational
Astronomy and RS holds Fairchild Foundation Senior Fellowship in
Theoretical Astrophysics.
 The VLA is a
facility of the National Science Foundation operated under cooperative
agreement by Associated Universities, Inc. 
The W. M. Keck Observatory is operated by the California Association
for Research in Astronomy, a scientific partnership among California
Institute of Technology, the University of California and the National
Aeronautics and Space Administration.  It was made possible by the
generous financial support of the W. M. Keck Foundation.}

\end{document}